\documentclass[a4paper,fleqn]{cas-dc}
\usepackage[numbers]{natbib}
\usepackage{hyperref}
\hypersetup{colorlinks=true,
linkcolor=blue,
anchorcolor=blue,
citecolor=blue}

\usepackage{multirow}
\usepackage{array} 
\usepackage{makecell}

\usepackage{cleveref}
\crefname{figure}{Fig.}{Figs.}
\crefname{table}{Table}{Tables}
\renewcommand{\figurename}{Fig.} 
\usepackage{graphicx}
\usepackage{caption}

\begin{document}
\begin{sloppypar}
\let\WriteBookmarks\relax
\def\floatpagepagefraction{1}
\def\textpagefraction{.001}
\let\printorcid\relax 

\shorttitle{}    

\shortauthors{Lanlan Kang et al.}

\title[mode = title]{Automated Quality Evaluation of Cervical Cytopathology Whole Slide Images Based on Content Analysis}

\author[1]{Lanlan Kang}\ead{kllsgdka@gmail.com}
\author[1]{Jian Wang}\ead{wangjianlydia@163.com}
\author[1,2]{Jian Qin}\ead{qinjian@ahut.edu.cn}
\cormark[1]
\author[1]{Yiqin Liang}\ead{2110402071@hrbust.com.cn}
\author[3]{Yongjun He}\ead{heyongjun@hit.edu.cn}

\address[1]{School of Computer Science and Technology, Harbin University of Science and Technology, Harbin, China} 
\address[2]{School of Computer Science and Technology, Anhui University of Technology, Maanshan, China} 
\address[3]{School of Computer Science and Technology, Harbin Institute of Technology, Harbin, China} 
\cortext[1]{Corresponding author.}  

\date{}

\begin{abstract}
    The ThinPrep Cytologic Test (TCT) is the most widely used method for cervical cancer screening, and the sample quality directly impacts the accuracy of the diagnosis. Traditional manual evaluation methods rely on the observation of pathologist under microscopes. These methods exhibit high subjectivity, high cost, long duration, and low reliability. With the development of computer-aided diagnosis (CAD), an automated quality assessment system that performs at the level of a professional pathologist is necessary. To address this need, we propose a fully automated quality assessment method for Cervical Cytopathology Whole Slide Images (WSIs) based on The Bethesda System (TBS) diagnostic standards, artificial intelligence algorithms, and the characteristics of clinical data. The method analysis the context of WSIs to quantify quality evaluation metrics which are focused by TBS such as staining quality, cell counts and cell mass proportion through multiple models including object detection, classification and segmentation. Subsequently, the XGBoost model is used to mine the attention paid by pathologists to different quality evaluation metrics when evaluating samples, thereby obtaining a comprehensive WSI sample score calculation model. Experimental results on 100 WSIs demonstrate that the proposed evaluation method has significant advantages in terms of speed and consistency.
\end{abstract}

\begin{keywords}
image quality evaluation \sep 
quality control \sep 
whole slide image \sep
digital pathology \sep
Computer-Aided Diagnosis
\end{keywords}
\maketitle
\section{Introduction}
Cervical cancer is one of the deadliest killers of women, claiming a large number of lives every year. Early detection and treatment are effective ways to deal with cervical cancer. TCT is the most widely used method for cervical cancer screening, and sample quality directly affects the accuracy and reliability of diagnostic results. Low-quality samples can lead to incorrect diagnosis \citep{ahmad2021artificial}. Therefore, it is essential to ensure the sample quality through assessment methods. Traditional sample quality evaluation relies on manual manipulation and observation by pathologists under microscopes, which is high subjectivity, high costs, long durations, and low reliability. In recent years, the advancement of pathology image digitization and artificial intelligence technology has led to a significant number of WSIs generated daily, posing new challenges to quality evaluation. In this context, conventional manual quality assessment methods can no longer effectively handle the quality assessment of large-scale WSIs. Assessment methods based on deep learning models offer a practical solution.
    \par Intelligent pathological image quality assessment aims to translate the evaluation methods and standards of pathology experts into algorithms that run on computers, achieving precise and efficient automated evaluation. Currently, some deep learning-based methods have been proposed. These methods compute visual quality metrics, such as sharpness, contrast, brightness \citep{albuquerque2021deep, shresthaQuantitativeApproachEvaluate2016a, yagiColorStandardizationOptimization2011, shresthaColorAccuracyReproducibility2014, chengAssessingColorReproducibility2013}, and occlusion factors like tissue folding and bubbles. Some research focuses on computing metrics of blur and noise \citep{shresthaQuantitativeApproachEvaluate2016a, hashimotoReferencelessImageQuality2012a, senarasDeepFocusDetectionOutoffocus2018b, choiNoreferenceImageQuality2009, hossainPracticalImageQuality2018a} with linear regression analysis using the sharpness and noise information from the training dataset. Others employ UNet networks \citep{ronneberger2015u} and various improved UNet architecture techniques to segment \citep{haghighat2022automated, hossainTissueArtifactSegmentation2023a,shakhawatPaperAutomaticQuality2020} and classify \citep{shakarami2023tcnn, kanwalVisionTransformersSmall2023b, ke2023artifact} quality regions, then perform severity assessments of image quality.
    \par However, these methods cannot meet the requirements of cervical cytopathology image quality assessment in practical applications. First, existing methods mostly focus on assessing histopathologic images and are not suitable for the quality assessment of TCT cytopathologic images. Second, existing methods often limited themselves to computing a single metric, locking a comprehensive evaluation of image quality. In addition, existing methods do not analyze image content, so they cannot compute metrics in TBS. According to TBS, a quality sample must contain at least 5,000 cells, as too few cells can result in a higher rate of missed diagnoses; additionally, impeding factors such as cell masses, markers, and air/gel bubbles can lead to misdiagnosis.
    \par To solve the above problem, we propose a comprehensive and quantitative evaluation method for cervical cytopathology WSI, considering the smear specimen quality grading standards outlined in TBS, the image quality requirements of computer-aided diagnosis systems, and the quality issues present in clinical data. Specifically, we not only compute conventional quality assessment metrics required by CAD, such as sharpness and noise but also utilize deep learning models to analyze the content of cervical cytopathology images. In this way, we can compute occlusion conditions for markers and neutrophil, interference factors for air/gel bubbles, and the standardized staining quality. We adopted a multi-model approach analysis by using multiple deep-learning models of object detection, image classification, image segmentation, and stain normalization. By integrating the prediction results of, deep learning models, we provide a comprehensive quality evaluation method. Our contributions are summarized as follows:
    \begin{enumerate}
        \item We design a comprehensive evaluation metric system based on TBS diagnostic standards and pathologists' evaluation method. This system includes general image quality metrics such as sharpness, and metrics of TBS, such as cell count and staining quality, and provides reliable quality assurance for subsequent diagnostic processes.
        \item Based on the proposed metrics, we propose a quality scoring method based on content analysis for cervical cytopathology WSI. This method leverages deep learning models of object detection, image classification, image segmentation, and stain normalization to compute the metrics and quality scores of WSIs. We achieve more accurate and consistent evaluations of image quality.
        \item We validated the proposed quality assessment method through experiments on a dataset of 100 cervical cytopathology WSIs. The results demonstrated that this method outperforms traditional ones in terms of efficiency, accuracy, and consistency.
    \end{enumerate}

\section{Related work}
    Existing research methods primarily focus on evaluating the sharpness and artifacts of tissue pathological images. Sharpness metrics are typically computed by analyzing the sharpness of edges and details in images. In contrast, artifact metrics address unnecessary or disruptive elements of images, such as tissue folds, bubbles, and markers. In this section, we will introduce related work of pathological image quality assessment.
    \par In natural images, quality issues usually arise from distortions or visual artifacts during the image acquisition or processing stages, primarily manifesting as problems with image sharpness. In the field of medical image evaluation, these concepts also play a referential role. Researchers have designed various general quality assessment methods based on deep learning technology \citep{liuRankiqaLearningRankings2017, bosseDeepNeuralNetworks2017, maEndendBlindImage2017}. Chen et al.  \citep{chen2019no} proposed a high-performance, no-reference image quality assessment method based on image entropy. This method can evaluate the quality of different types of distorted images with low complexity. Hosseini et al. \citep{hosseini2019encoding} propose a novel design of human visual system response in a convolutional filter form to decompose meaningful features that are closely tied to image sharpness level. Kim et al. \citep{kim2017deep} assay progress in this rapidly evolving field, focusing, in particular, on recent CNN-based picture-quality prediction models that deliver excellent results in a large, real-world, picture-quality database. Zhang et al. \citep{2019Blind} proposed a deep bilinear model for no-reference image quality assessment, achieving unified quality evaluation. For the quality assessment of pathological WSIs, it is essential to consider not only the requirements of CAD systems but also the requirements of TCT, such as epithelial cell number, tissue texture, staining quality, and interference factors.
    \par In recent years, the development of deep learning technology has provided new approaches for WSI image quality assessment. Some researchers have borrowed methods from the field of natural image processing to assess the quality of entire WSIs based on sharpness metrics. Hosseini et al. \citep{hosseiniFocusQualityAssessment2019} proposed no-reference focus quality assessment metrics for digital pathology images, quantifying patch-level focus quality through these metrics. Senaras8 et al. developed a deep learning-based software called, DeepFocus, which can automatically detect and segment blurry areas in whole slide images. Wang et al. \citep{wangFocuslitennHighEfficiency2020a} designed a first-of-its-kind blind IQA model for multiplying distorted visual content based on a deep end-to-end convolutional neural network. Geng et al. \citep{gengCervicalCytopathologyImage2022a} proposed a refocusing method for cervical cytopathology images using multi-scale attention features and domain normalization, contributing to improved performance in subsequent analysis tasks. These methods have achieved certain successes in out-of-focus quality assessment but are limited to the blurred areas images.
    \par Some researchers have also considered detecting tissue artifacts (such as markers, bubbles, and tissue folds) to assess WSI quality. Shakhawat et al. \citep{shakhawatPaperAutomaticQuality2020} propose a method for a practical system to assess WSI quality by distinguishing the origins of quality degradation. In the method, artifacts are first detected by a support vector machine, and then the quality of regions excluding artifacts is evaluated. Kanwal et al. \citep{kanwalVisionTransformersSmall2023b} proposed a student-teacher method to enhance the classification performance of ViT in bubble detection tasks, thereby improving artifact detection. Shakarami et al. \citep{shakarami2023tcnn} proposed a new method for detecting artifacts based on a transformer-based convolutional neural network, achieving improved performance.
    \par Additionally, researchers have attempted to evaluate WSI quality from other perspectives. Kanwal et al. \citep{kanwalDevilDetailsWhole2022a} demonstrated that WSI preprocessing can significantly improve the performance of computational pathology systems, and conducted detailed studies on evaluation techniques in three different areas: artifact detection, color variation, and pathology data augmentation. Janowczyk et al. \citep{janowczyk2019histoqc} developed an open-source tool for rapid quality control, which uses an interactive interface for real-time visualization and filtering to assist users in artifact detection. In Shrestha et al.'s study \citep{shresthaQuantitativeApproachEvaluate2016a} , a quantitative measurement method for WSIs was proposed to evaluate by computing sharpness, contrast, brightness, uniform illumination, and color separation. The goal was to assess image quality consistency while controlling the uniformity of WSI quality across multiple scanners. Haghighat et al. \citep{haghighat2021pathprofiler} trained a multitask neural network for quality control of large prostate WSI datasets, assessing quality based on overall usability, focus, and H\&E staining quality of the WSIs. Ke et al. \citep{ke2023artifact} proposed an AR-Classifier artifact recognition network that distinguish common artifacts from normal tissue, such as tissue folds, marking dye, tattoo pigment, spots, and out-of-focus areas, for automated quality control.
    \par We can see that researchers are continuously advancing pathological WSI quality assessment technologies to improve their accuracy and practicality. In traditional cervical cancer screening, pathologists rely on visual inspection to observe the thickness and area of the specimen smear, as well as the number and coverage of squamous epithelial cells under the microscope. Additionally, current research mainly focuses on the field of histopathology, which usually only considers a single quality metric and lacks a comprehensive analysis of the image content, and thus fails to meet the image quality requirements in practical application scenarios. To address this problem, we designed a quality assessment approach for cervical cytopathology WSIs by thoroughly analyzing the content of cervical cytopathology images and basing it on TBS diagnostic standards.
\section{methods}
\begin{figure*}[t]
    \centering
    \includegraphics[width=0.9\linewidth]{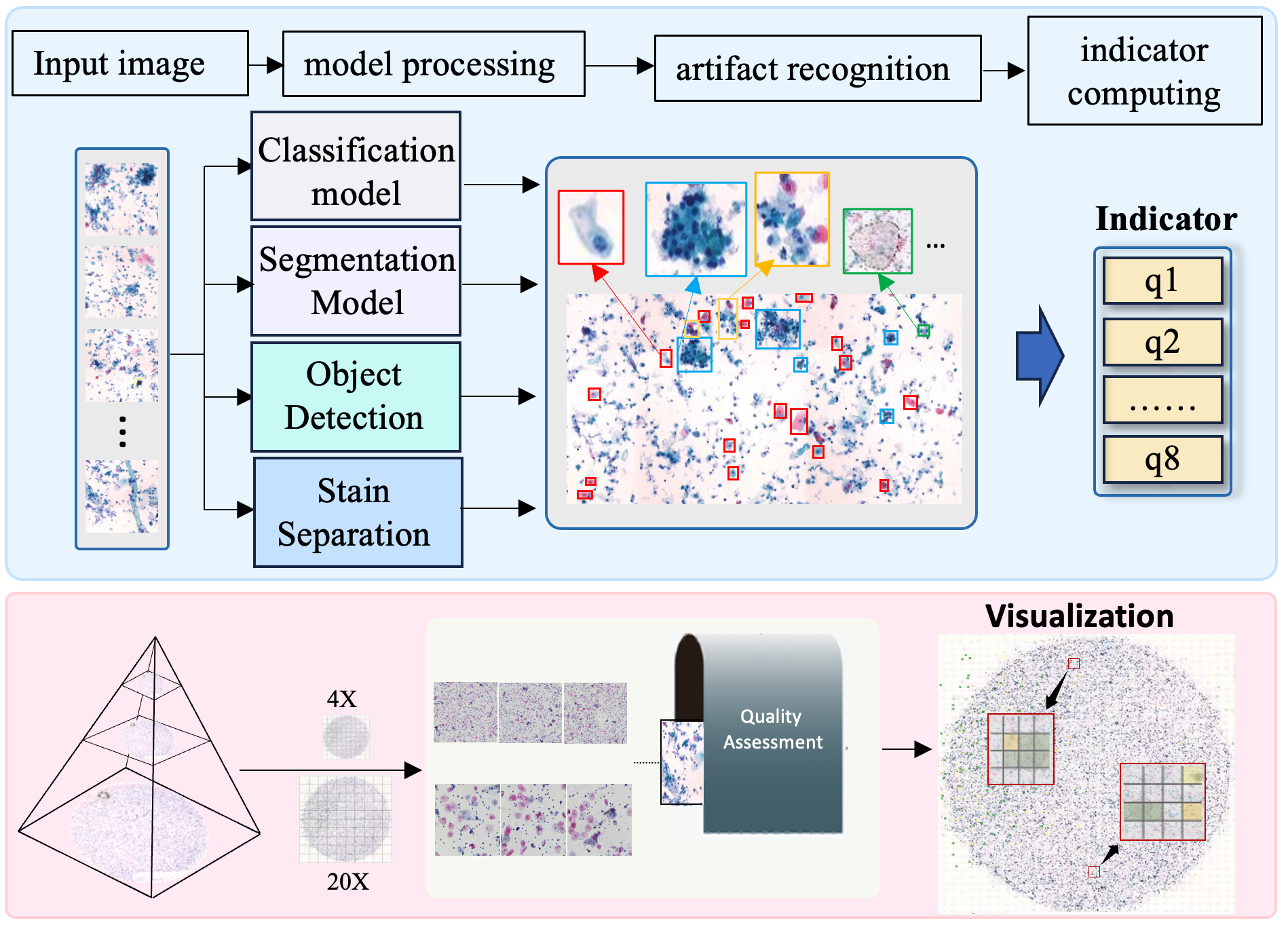}
    \caption{Pipeline of the proposed method.}
    \label{Fig.1}
    \end{figure*}

    To comprehensively and accurately describe WSI quality, we propose multiple evaluation metrics that constitute a comprehensive metric system. The metric system includes grid-like imaging, air/gel bubbles, markers, cell count, cell masses, neutrophils, substandard staining, and out-of-focus regions. A summary of these metrics are presented in \hyperref[table:1]{\tablename~\ref{table:1}}. The quality evaluation pipeline is shown in \hyperref[Fig.1]{\figurename~\ref{Fig.1}}. This method selects the appropriate image magnification for detection based on the characteristics of the quality problem to ensure the accuracy and validity of the detection. Firstly, the WSIs are cut into patches. Next, several models are utilized to compute metrics. Finally, the computed metrics are integrated by XGBoost model to obtain the WSI comprehensive quality score. Below we describe the calculation of each metric specifically.

    \begin{footnotesize}
    \begin{table}[t] 
    \caption{
    \label{table:1}Overview of the interpretation and methodological realization of the metrics.} 
    \begin{tabular*}{\tblwidth}{@{}LLc@{}}
    \toprule Metrics & Description of metrics & Quantitative descriptions \\ 
    \midrule q1 & Grid-like Imaging & \multirow{8}{*}{\makecell{[0,1] \\The lower the value the \\higher the degree.}}\\ 
    q2 & Out of Focus               & ~\\
    q3 & Marker                     & ~\\ 
    q4 & Air/Gel Bubble             & ~\\ 
    q5 & Staining standard         & ~\\ 
    q6 & Cell                      & ~\\ 
    q7 & Cell Mass                  & ~\\ 
    q8 & Neutrophil                 & ~\\ 
    
    \bottomrule 
    \end{tabular*}
    \end{table}
    \end{footnotesize}

\subsection{Grid-like imaging metric}
    A WSI is created by stitching together multiple scans of small area images. In the process of image scanning, if there is uneven illumination, scanner failure, or improper parameter setting, grids will be generated at the image stitching points, which will affect image quality. When the illumination is uniform, the pixel gray values of content-free images should be in a homogeneous state. For images with grid-like imaging, the differences in gray values are relatively large. Therefore, to evaluate grid-like artifacts in WSI, we calculate the mean and variance of five image patches without content in WSI. The mean, variance and $q1$ calculation can be represented by the following formula:
    \begin{equation}
        {{M}_{patch}}=\frac{1}{N}\sum\limits_{i=1}^{1}{{{I}_{i}}}
    \end{equation}
    \begin{equation}
        {{V}_{patch}}=\frac{1}{N}\sum\limits_{i=1}^{1}{\left( {{I}_{i}}-{{M}_{patch}} \right)}
    \end{equation}
    \begin{equation}
        {{V}_{wsi}}=\frac{1}{5}\sum\limits_{i=1}^{5}{{{V}_{patch}}_{i}}
    \end{equation}
    \begin{equation}
        q1=\frac{\left| {{V}_{wsi}}-{{V}_{nogrid}} \right|}{{{V}_{nogrid}}}
    \end{equation}
    where ${{M}_{patch}}$ is the mean of  pixel value in the patch, ${{V}_{patch}}$ is the variance of pixel value in the patch, ${{I}_{i}}$ is the $i$-th pixel value of patch, and $N$ is the total number of pixels in the patch. ${{V}_{wsi}}$ is the variance of pixel value of five patches without content in WSI. ${{V}_{nogrid}}$ is the variance of pixel values in patch without gird-like.
\subsection{Out-of-Focus metric}
    During the scanning process, focus failure can cause image blurriness, making it difficult for physicians to diagnose disease from the images. We use a sharpness evaluation model to detect out-of-focus problems. Focus information is primarily encoded in edge information and images with high clarity have sharper and more distinct edges. We can evaluate the sharpness of images by analyzing features such as edge strength, continuity, and clarity. Therefore, we use shallow deep learning model to detect sharpness of images. We build a simple model, called FocusAttNet, which includes only a convolutional layer and an attention layer. Our approach aims to reduce the layer complication in deep learning for sharpness evaluation, while still being able to the high performance for the task.
    \par Assuming that the sharpness level of sufficiently small mosaic patches $X\in {{R}^{HxWx3}}$ extracted from the WSI is uniform. We first convolve the patch with a set of kernels $\theta \in {{R}^{hxwx3xN}}$ and then apply a nonlinear pooling function $y$ to predict the sharpness of the patch:
    \begin{equation}
        y={{p}_{NL}}\left( \sum\limits_{k=1}^{3}{\phi *{{X}_{k}}+b} \right)
    \end{equation}
    where ${{\phi }_{k}}\in {{\mathbb{R}}^{h\times w\times N}}$ is the convolution kernel for the $k$-th input channel. ${{X}_{k}}\in {{\mathbb{R}}^{H\times W}}$ represents the $k$-th channel of the input image patch, $b\in {{\mathbb{R}}^{N}}$ is the bias term, and $y\in \mathbb{R}$ is the predicted score for $X$. The 2D convolution operator $*$ has a stride of 5, The function ${{p}_{NL}}$ is a nonlinear pooling function that maps the 2D response to the overall sharpness score $y$. All kernel sizes are set to $h=w=7$. The use of the pooling function ${{p}_{NL}}$ produces nonlinearity in the model, significantly enhancing the approximation capability of the simple model. It is defined as follows:
    \begin{equation}
        {{p}_{NL}}\left( x \right)={{w}_{1}}\cdot \min \left( x \right)+{{w}_{2}}\cdot \max \left( x \right)+{{w}_{3}}
    \end{equation}
    where $x\in {{\mathbb{R}}^{\frac{H-h+7}{5}\times \frac{W-w+7}{5}\times N}}$ is the response produced by the convolution, and ${{w}_{1}}\in {{\mathbb{R}}^{N}}$, ${{w}_{2}}\in {{\mathbb{R}}^{N}}$, ${{w}_{3}}\in {{\mathbb{R}}}$ are the trainable parameters.
    \par Through the above calculations, we obtain the patch-level sharpness. We set the stride to 128x128 pixels for dense sampling across the entire patch, using the average score ${{q}_{patch}}$ as the sharpness of the patch. We then compute the WSI sharpness by averaging the patch sharpness scores, with predicted scores ranging from 0 to 12. To standardize the scores and reflect image blurriness, we normalize this range to between 0 and 1. The patches used for this process are obtained through interval sampling of the WSI. As shown in the following formula:
    \begin{equation}
        {{q}_{patch}}=\frac{1}{n}\sum\limits_{i=1}^{n}{{{y}_{i}}}
    \end{equation}
    \begin{equation}
        q2=\frac{1}{m}\sum\limits_{i=1}^{m}{{{q}_{\text{patch}i}}}
    \end{equation}
    where ${{y}_{i}}$ represents the predicted score for the $i$-th sampled patch, $i\in \left[ 1n \right]$, $n$ is the number of densely sampled small patches. $q2$ is the sharpness metric for the WSI, and $m$ denotes the total number of patches.
\subsection{Marker, air/gel bubble metric}
\begin{figure}[h]
    \centering
    \includegraphics[width=0.9\linewidth]{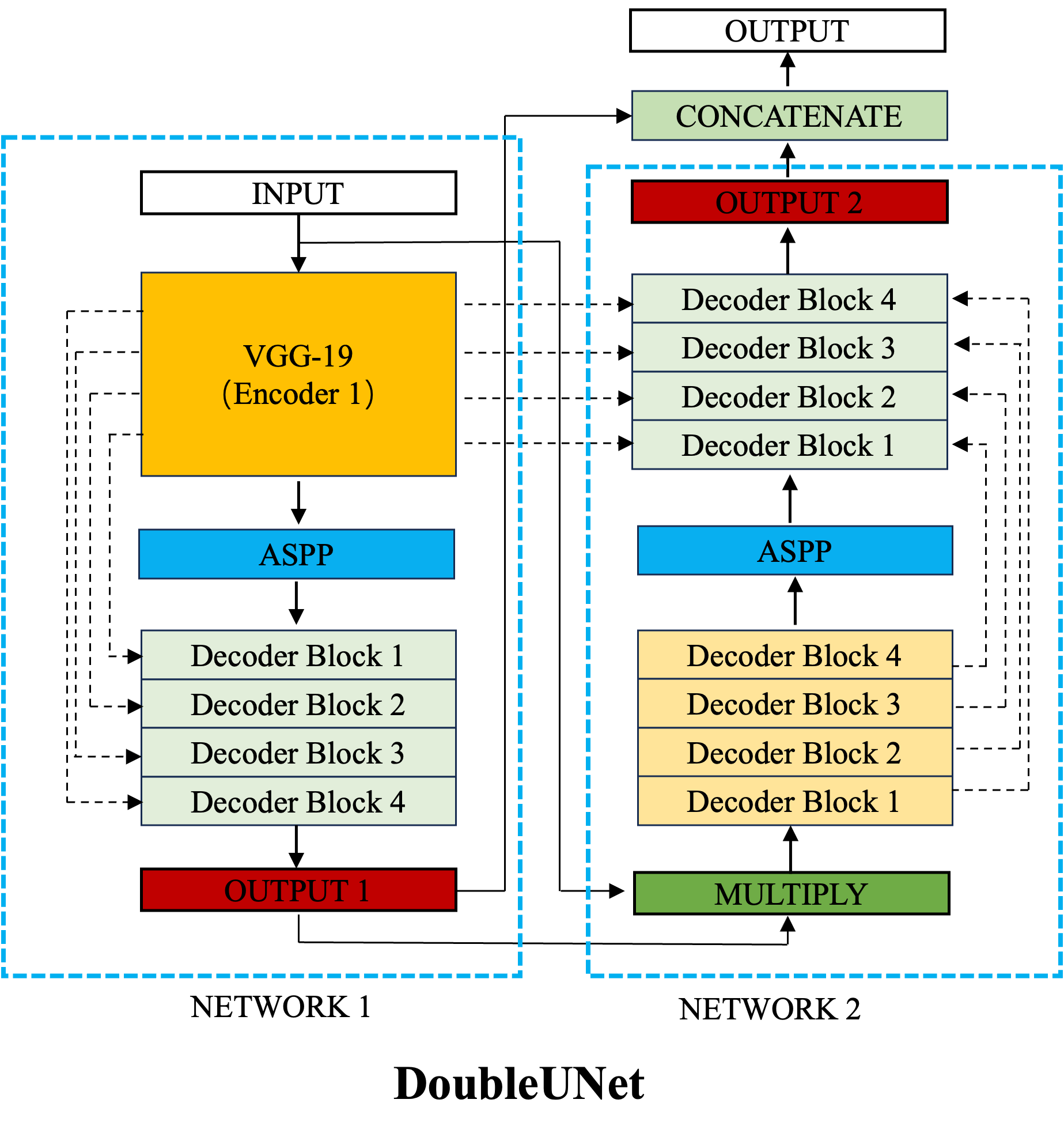}
    \caption{Model for Segmenting Marker and Air/gel Bubbles.}
    \label{Fig.2}
    \end{figure}

    DoubleUNet has recently shown excellent performance in medical image segmentation tasks. For the detection of marker and air/gel bubble areas in WSIs, we use the DoubleUnet network. DoubleUnet architecture uses the VGG-19 model as the encoder. The first layer of UNet generates an output which is multiplied with the primary input and that multiplied output is given as input to the second layer of UNet. The primary output and the secondary output concatenate and generate the final output. As illustrated in \hyperref[Fig.2]{\figurename~\ref{Fig.2}}.
    \par We compute the area ratios of markers and air/gel bubbles for each patch and assign corresponding quality scores $q{{3}_{i}}$ and $q{{4}_{i}}$ within the range $\left [ 0,1 \right ]$, where 1 indicates best quality and 0 represents the worst quality.These patch-specific quality scores are combined using a weighted average to evaluate the markers and air/gel bubbles across the entire WSI. The calculation formulas are as follows:
    \begin{equation}
        q{{3}_{i}}=\frac{{{A}_{mar\ker }}}{{{A}_{patch}}}
    \end{equation}
    \begin{equation}
        q3=\frac{1}{n}\sum\limits_{i=1}^{n}{q{{3}_{i}}}
    \end{equation}
    where $q{{3}_{i}}$ is the marker metric for the $i$-th patch, $i\in \left[ 1,n \right]$, ${{A}_{marker}}$ is the area of marker in the patch, ${{A}_{patch}}$ is the area of the patch, and $n$ is the number of patches containing markers. Similarly, $q4$ is the air/gel bubble metric.
\subsection{Staining standard metric}
\begin{figure}[t]
    \centering
    \includegraphics[width=0.9\linewidth]{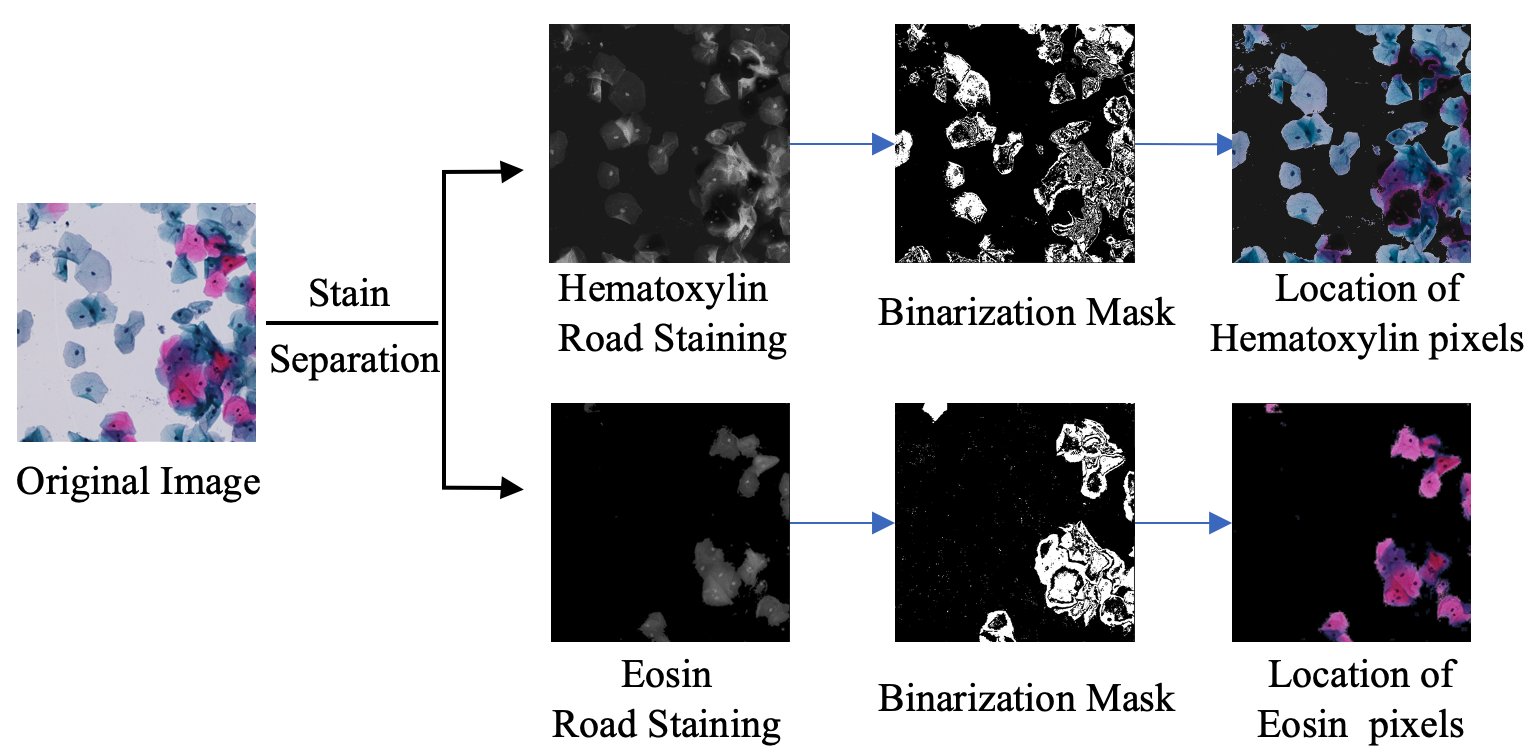}
    \caption{Cervical Cell Staining Separation Process.}
    \label{Fig.3}
    \end{figure}

    In cytopathological examinations, the Papanicolaou stain is one of the most commonly used staining methods for exfoliated cells. The Papanicolaou staining solution mainly consists of hematoxylin, eosin, orange G, EA50, etc. Pathologists analyze cellular information in images for clinical diagnosis. Therefore, it's crucial to make staining sharpness and uniformity for abnormal cell detection. This study employs an adaptive color deconvolution method to separate the staining components in the image for the purpose of evaluating whether the sample staining meets the requirements. The sample data is stained with hematoxylin and eosin (H\&E). Subsequently, threshold segmentation is applied to the separated images to obtain masks for the H\&E channels. By performing an “AND” operation between these masks and the original image, the positions of H\&E-stained pixels are identified. Then the average gray value ${{V}_{gray}}$ of these pixels are calculated. The staining separation process is illustrated in \hyperref[Fig.3]{\figurename~\ref{Fig.3}}.
    \par To determine the Staining standard for H\&E, this study conducted a clustering analysis on standard samples provided by doctors to establish the optimal staining intensity range ${V}_{stdgray\min}$ $\sim $  ${V}_{stdgray\max }$ for the eosin channel and the hematoxylin channel. $q5$ can be represented by the following formula:
    {\small
    \begin{equation}
    q5=\begin{cases}
    1-\frac{V_{gray}-V_{stdgraymax}}{V_{stdgraymax}} & \text{ if } \frac{V_{gray}}{V_{stdgraymax}}\leqslant 2 \\
    1 & \text{ if } V_{stdgraymin}\leqslant V_{gray} \leqslant V_{stdgraymax} \\
    1-\frac{V_{stdgraymin}-V_{gray}}{V_{stdgraymin}} & \text{ if } \frac{V_{gray}}{V_{stdgraymin}}\leqslant 1 \\
    0 & \text{ other }  
    \end{cases}
    \end{equation}
    }

\subsection{Squamous epithelial cell, cell mass, and neutrophil metrics}
\begin{figure}[t]
    \centering
    \includegraphics[width=0.9\linewidth]{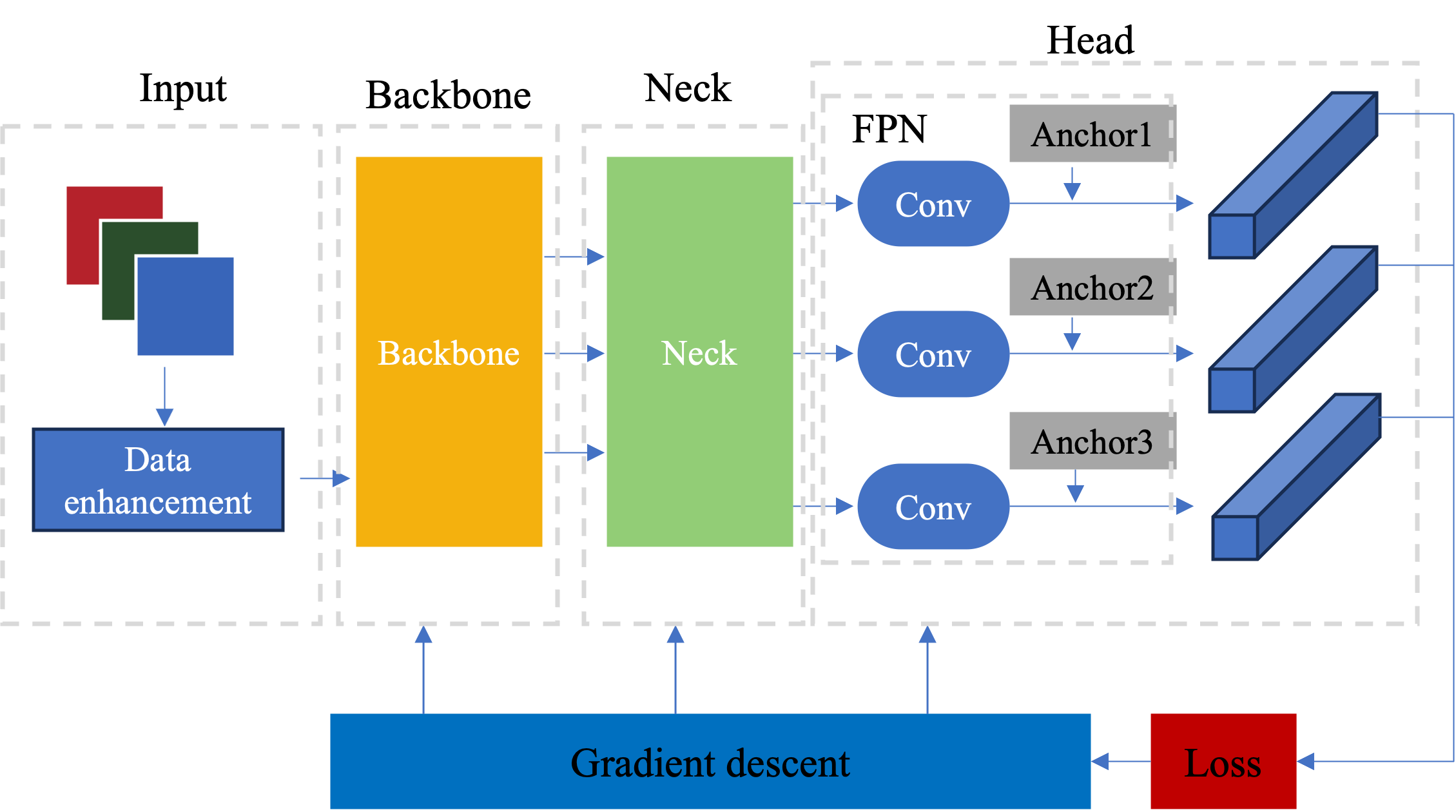}
    \caption{Network structure of YOLOV5.}
    \label{Fig.4}
    \end{figure}

    According to the TBS diagnostic criteria for cervical cytopathology, a satisfied cervical cytopathology specimen should contain between 5000 and 10000 squamous epithelial cells, fewer than 50 cell masses, and less than 25\% of epithelial cells obscured by neutrophils. If 50\% to 75\% of the epithelial cells are obscured, the quality metrics section of the report should note inflammatory obscuration. If more than 75\% of the epithelial cells are obscured, then the sample should be regarded as unsatisfactory. We use the YOLOv5 model to calculate quality evaluation metrics for cervical cytopathology images, as shown in \hyperref[Fig.4]{\figurename~\ref{Fig.4}}. Based on the results, we count squamous epithelial cells and use threshold segmentation to calculate the area proportions of cell masses and neutrophils.
    \par Based on the results detected by the model and the quality standards, we calculate the metrics for the number of squamous epithelial cells (metric q${{7}}$), the proportion of cell masses (metric q${{8}}$), and the neutrophil obscuration (metric q${{9}}$). The calculation formulas are as follows:
    
    \begin{equation}
        q6=\begin{cases}
        \frac{x}{5000} & \text{ if } x< 5000 \\
        1 & \text{ if } x\geq 5000 
        \end{cases}
    \end{equation}

    \begin{equation}
        q7=\begin{cases}
        1 & \text{ if } x\leqslant  50 \\
        \frac{50}{x} & \text{ if } x>  50 
        \end{cases}
    \end{equation}
    
    \begin{equation}
        q8=1-\frac{{{S}_{neutrophil}}}{S}
    \end{equation}
    where $x$ is the number of squamous epithelial cells, ${S}$ the effective field of view area of WSI, and ${{S}_{neutrophil}}$ is the area of neutrophils.
\subsection{Quality score of WSI}
    Since a single metric cannot detect all quality issues comprehensively, we employ an ensemble learning approach. Specifically, we perform strategy fusion on the metrics obtained from multiple models based on patches to calculate the quality score for the WSI. To emulate pathologists' scoring method, we had them annotate the data and assign a 0-10 score to each WSI. We first quantified and normalized various metrics based on predefined rules and model confidence levels. These normalized metrics then served as input for an XGBoost model, using the pathologists' ratings as learning labels to train the model to accurately replicate the pathologists' scoring methodology.
    \par Typical WSI images are in the order of gigapixels and are usually stored in multi-resolution pyramidal format. At 4x magnification, the WSI is divided into non-overlapping patches with size of 512x512. Each patch needs to be preprocessed to exclude patches with predominantly white pixels (patches where more than 80\% of the pixels exceed 200 are considered white and excluded). For the remaining patches, grid-like imaging, marker, and air/gel bubble metrics are calculated using trained models designed to detect specific artifacts in these patches. At 20x magnification, the WSI is again divided into non-overlapping patches of size 512x512, with the same preprocessing to exclude white patches. Relevant models are then applied to these patches to calculate out-of-focus metric, squamous epithelial cell count metrics, cell mass proportion metric, and neutrophil coverage metric.
    \par XGBoost, short for eXtreme Gradient Boosting, is a robust ensemble learning technique. Its core components are decision trees, which collectively constitute the XGBoost model. For n instances and m features, the formula is given by:
    \begin{equation}
        D=\left\{ \left( {{x}_{i}},{{y}_{i}} \right) \right\}\left( \left| D \right|=n,{{x}_{i}}\in {{\mathbb{R}}^{m}},{{y}_{i}}\in \mathbb{R} \right)
    \end{equation}
    \begin{equation}
        \widehat{{{y}_{i}}}=\sum\limits_{i=2}^{k}{{{f}_{t}}\left( {{x}_{i}} \right)},{{f}_{t}}\in F
    \end{equation}
    where ${{f}_{k}}$ is the $k$-th decision tree. $\widehat{{{y}_{i}}}$ is the predicted value for the $i$-th sample, $F$ is the function space composed of $k$ base functions, ${{x}_{i}}$ is the $i$-th extracted feature metric.
    \par Based on WSI metric scores and the overall quality score, we assess the suitability of cervical WSI for downstream diagnostic tasks. WSI with scores above 6 are archived for future research and testing. Lower-scoring WSI are addressed by re-preparing slides or re-scanning, depending on whether the issues arise from preparation or scanning.
\section{Experiments and analysis}
\subsection{Dataset}
\begin{figure*}[h]
    \centering
    \includegraphics[width=0.9\linewidth]{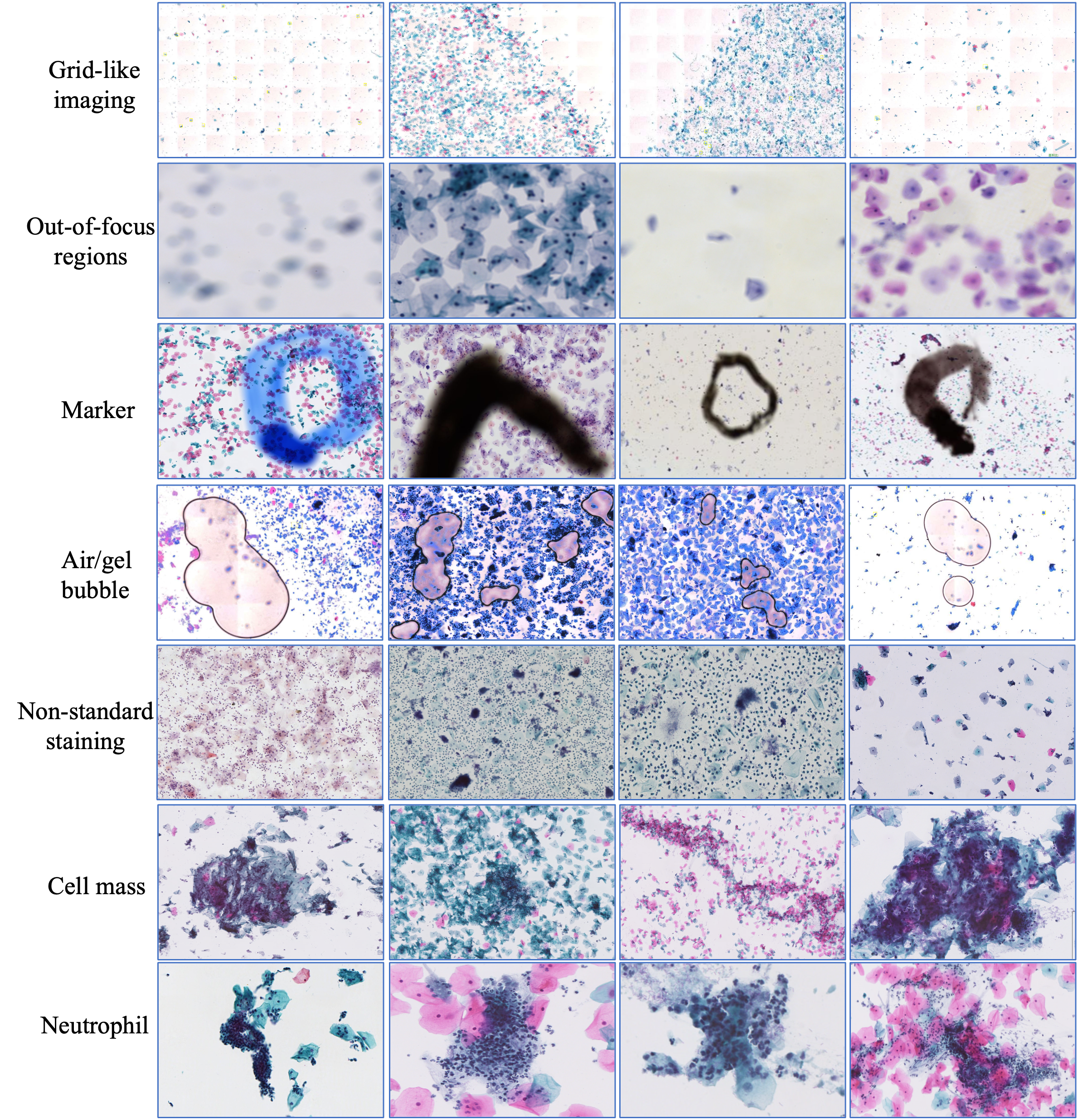}
    \caption{Examples of quality issues in cervical cytopathology images.}
    \label{Fig.5}
    \end{figure*}

\begin{figure*}[t]
    \centering
    \includegraphics[width=0.8\linewidth,height=0.5\linewidth]{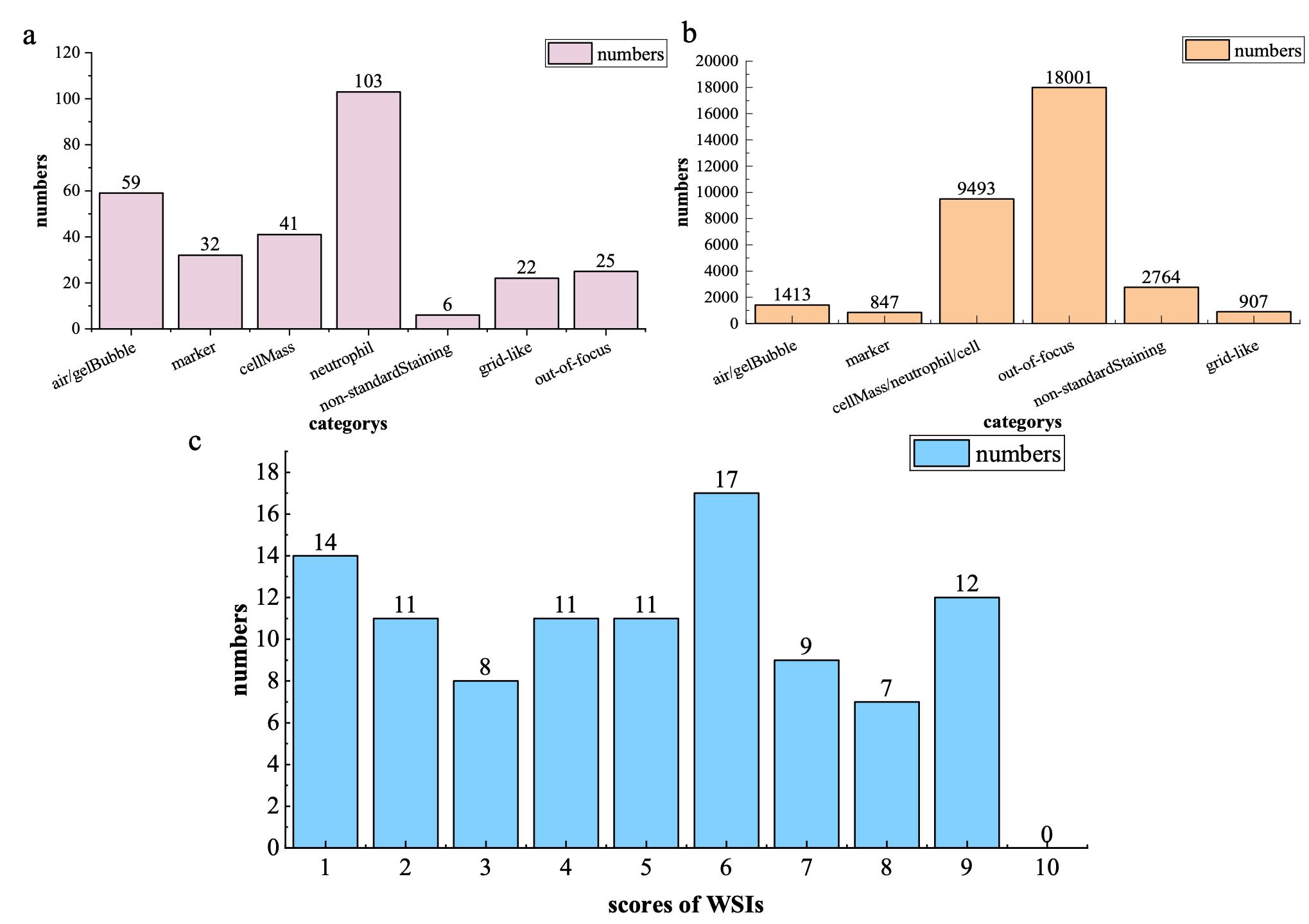}
    \caption{Dataset Details.}
    \label{Fig.6}
    \end{figure*}

    We evaluate our method on a dataset with 302 WSIs. The data includes eight types of quality issues: air/gel bubble, marker, cell mass, neutrophils, grid-like imaging, substandard staining, out-of-focus and insufficient cell number, as illustrated in \hyperref[Fig.5]{\figurename~\ref{Fig.5}}. The number of WSI for each type of quality issue is shown in \hyperref[Fig.6]{\figurename~\ref{Fig.6}}a. The slides were fixed in formalin and stained with Papanicolaou stain, and they were stored in NDPI, SVS, and TIFF formats with dimensions of approximately 80,000 x 80,000 pixels. All WSIs are anonymized and comply with ethical standards. 
    \par To train the model for calculating the metrics, we collect several image patches with various resolution of 256x256, 512x512, and 1024x1024. The training set for metrics calculation contains 33425 patches. The specific distribution of the dataset in \hyperref[Fig.6]{\figurename~\ref{Fig.6}}b, and the WSI scores provided by pathologists are also shown in \hyperref[Fig.6]{\figurename~\ref{Fig.6}}c.

\subsection{Implementation details}
    We implement the proposed model in PyTorch framework. During training, since the model cannot directly process the entire WSI, we further partition it into smaller patches of sizes 256x256, 512x512, and 1024x1024. The experiments are organized into three main sections: sharpness assessment, artifact segmentation, and object detection. We identified the optimal configurations through experimentation. The hyperparameters for the models used are detailed in Table \ref{table:2}.

    \begin{tiny}
    \begin{table}[t] 
    \caption{
    \label{table:2}Hyperparameters for the models used in training.} 
    \begin{tabular*}{\tblwidth}{@{}LLL@{}}
    \toprule Model     & Parameter & Value \\ 
    \midrule 
    \multirow{5}{*}{\makecell{Sharpness\\Assessment\\Model}}     & Input patch size & 1024x1024x3\\ 
    ~                   & Learn Rate       & Starting from 0.01\\
    ~                        & Optimizer        &  Adam\\ 
    ~                             & Epoch            &  120\\ 
    ~                             & Loss function    &  PLCC\\ 
    \midrule 
    \multirow{12}{*}{\makecell{Segmentation\\Model}} & Input patch size & 256x256x3\\
    ~ & \multirow{2}{*}{Optimizer} & \multirow{2}{*}{\makecell{\begin{tabular}[t]{@{}l@{}}Adam, Adamax,\\ \raggedright RMSprop\end{tabular}}}\\
    ~ & ~ & ~\\
    ~ & Learning Rate & Starting form 1e-4\\
    ~ & \multirow{3}{*}{ReduceLROnPlateau} & \multirow{3}{*}{\makecell[l]{monitor=val\_loss, \\ factor=0.1, \\patience=4}}\\
    ~ & ~ & ~\\
    ~ & ~ & ~\\
    ~ & \multirow{3}{*}{EarlyStopping} & \multirow{3}{*}{\makecell{\begin{tabular}[t]{@{}l@{}}monitor=val\_loss,\\ \raggedright patience=10\end{tabular}}}\\
    ~ & ~ & ~\\
    ~ & ~ & ~\\
    ~ & \multirow{2}{*}{Loss function} & \multirow{2}{*}{\makecell{\begin{tabular}[t]{@{}l@{}}Dice Coef Loss,\\ \raggedright binary cross entropy\end{tabular}}}\\
    ~ & ~ & ~\\
    \midrule
    \multirow{5}{*}{\makecell{Object\\Detection\\Model}} & \multirow{3}{*}{Learning Rate} & \multirow{3}{*}{\makecell[l]{Start from 0.01,\\momentum=0.871,\\weight decay=0.05}}\\
    ~ & ~ & ~\\
    ~ & ~ & ~\\
    ~ & Max number of iterations & 10000\\
    ~ & Epoch & 100\\

    \bottomrule 
    \end{tabular*}
    \end{table}
    \end{tiny}
    
\subsection{Evaluation metric of model}
    The metrics used for evaluation include accuracy, sensitivity, specificity, Spearman Rank Correlation Coefficient(SRCC), Pearson Linear Correlation Coefficient(PLCC), the receiver operating characteristic (ROC) curve, and the area under the ROC curve (AUC). The formulas for calculating these metrics are as follows:
    \begin{equation}
        Accuracy=\frac{TP+TN}{TP+TN+FP+FN}
    \end{equation}
    \begin{equation}
       Sensitivity=\frac{TP}{TP+FN}
    \end{equation}
    \begin{equation}
        Specificity=\frac{TN}{TN+FP}
    \end{equation}
    where $TP,TN,FP,FN$ represented the number of true positive, true negative, false positive, false negative. For the evaluation of the object detection model, we used commonly employed metrics in the field of object detection: precision, recall, and $mA{{P}_{0.5}}$.
    \begin{equation}
        {{d}_{i}}={{r}_{xi}}-{{r}_{yi}}
    \end{equation}
    \begin{equation}
        SRCC=1-\frac{6\sum\nolimits_{i=1}^{n}{{{d}_{i}}^{2}}}{n\left( {{n}^{2}}-1 \right)}
    \end{equation}
    \begin{equation}
        PLCC=\frac{\sum\limits_{i=1}^{n}{\left( {{x}_{i}}-\overline{x} \right)\left( {{y}_{i}}-\overline{y} \right)}}{\sqrt{\sum\limits_{i=1}^{n}{{{\left( {{x}_{i}}-\overline{x} \right)}^{2}}}}\sqrt{\sum\limits_{i=1}^{n}{{{\left( {{y}_{i}}-\overline{y} \right)}^{2}}}}}
    \end{equation}
    where $x=\left\{ {{x}_{1}},{{x}_{2}}\cdots ,{{x}_{n}} \right\}$ and $y=\left\{ {{y}_{1}},{{y}_{2}}\cdots ,{{y}_{n}} \right\}$. $n$ is the length of the data. ${{x}_{r}}=\left\{ {{r}_{{{x}_{1}}}},{{r}_{{{x}_{2}}}},\cdots ,{{r}_{{{x}_{n}}}} \right\}$ and ${{y}_{r}}=\left\{ {{r}_{{{y}_{1}}}},{{r}_{{{y}_{2}}}},\cdots ,{{r}_{{{y}_{n}}}} \right\}$ are obtained by sorting $x$ and $y$ in ascending order. ${{d}_{i}}$ is rank difference. $\overline{x}$ and $\overline{y}$ are the mean of $x$ and $y$.
\subsection{Results and analysis}
    To verify the effectiveness of each model in the evaluation method, we perform comparative experiments. 
    \par In the calculation of grid-like imaging metrics, we divide the WSI into multiple patches and compute the mean gray value and the standard deviation of contrast. Then we compare these with a threshold that is calculated by 689 no-grid-like patches. The experimental results indicate that the method can effectively detects grid-like imaging artifacts in the images. For the calculation of the threshold, first, we collect image samples without grid-like artifacts and calculate the mean gray value and variance standard deviation of their patches. Then, we perform a statistical analysis on these standard deviations to determine an appropriate threshold, typically selecting either the upper limit of the standard deviation or a value within the 95\% confidence interval.
    \par For the Sharpness model, we compared our proposed approach with existing methods, selecting FocusLiteNN, EONSS, DenseNet-13, and ResNet-10. We evaluated the selected Sharpness assessment models in terms of statistical correlation and classification performance as well as computational complexity on FocusPath and the cervical cytopathology datasets. We detailed the training configurations and hyperparameters in \hyperref[table:2]{\tablename~\ref{table:2}}. Evaluation metrics of model include Spearman's Rank Correlation Coefficient (SRCC), Pearson Linear Correlation Coefficient (PLCC), Area Under the Receiver Operating Characteristic Curve (ROC AUC), and Area Under the Precision-Recall Curve (PR AUC). We present the experimental results in \hyperref[table:3]{\tablename~\ref{table:3}}.
   
    \begin{small}
    \centering
    \begin{table}[t] 
    \caption{
    \label{table:3}SRCC, PLCC, ROC-AUC, PR-AUC Performance of 5 FQA Models on FocusPath Dataset and Cervical Cytopathology Datase(*).} 
    \begin{tabular}{>{\centering\arraybackslash}p{2.4cm} p{0.7cm} p{0.7cm} p{0.7cm} p{0.7cm} p{0.7cm}} 
    \toprule  \multirow{2}{*}{Model} & \multicolumn{2}{c}{FocusPath} & \multicolumn{2}{c}{Data*} & \multirow{2}{*}{Time/s}\\ 
    \cline{2-5}
    ~ & SRCC & PLCC &ROC & PR & ~\\
    \midrule 
    Our method & 0.8765 &	0.8627 &	0.9310 &	0.8499	& 0.034\\
    EONSS \citep{wangBlindQualityAssessment2019a} & 0.8374 & 0.8227 & 0.6542 & 0.6954 &	0.065\\
    FocusLiteNN \citep{wangFocuslitennHighEfficiency2020a}  & 0.8766 & 0.8668 & 0.6650 & 0.7022 & 0.043\\
    DenseNet-13 \citep{huangDenselyConnectedConvolutional2017} &  0.8974 & 	0.8898 & 	0.9490	 & 0.9477 & 	0.355\\
    ResNet-10 \citep{heDeepResidualLearning2016a} &  0.8507	 & 0.8387	 & 0.8566	 & 0.8489	 & 0.334\\
    \bottomrule 
    \end{tabular}
    \end{table}
    \end{small}

    \par DenseNet-13 achieved the best performance on both the FocusPath and cervical cytopathology datasets, which is attributed to its deep architecture and rich parameter configuration. However, the complexity of this model results in significant computational costs, limiting its practical feasibility. Our proposed improvement method significantly reduces computational time while maintaining performance comparable to DenseNet-13, achieving a 96.8\% acceleration. This indicates that by optimizing model structure and parameter configuration, substantial improvements in evaluation efficiency can be made without compromising accuracy. This optimization not only reduces computational resource consumption but also enhances the model's deployability in real clinical settings.
    \par
\begin{figure}[t]
    \centering
    \includegraphics[width=0.9\linewidth]{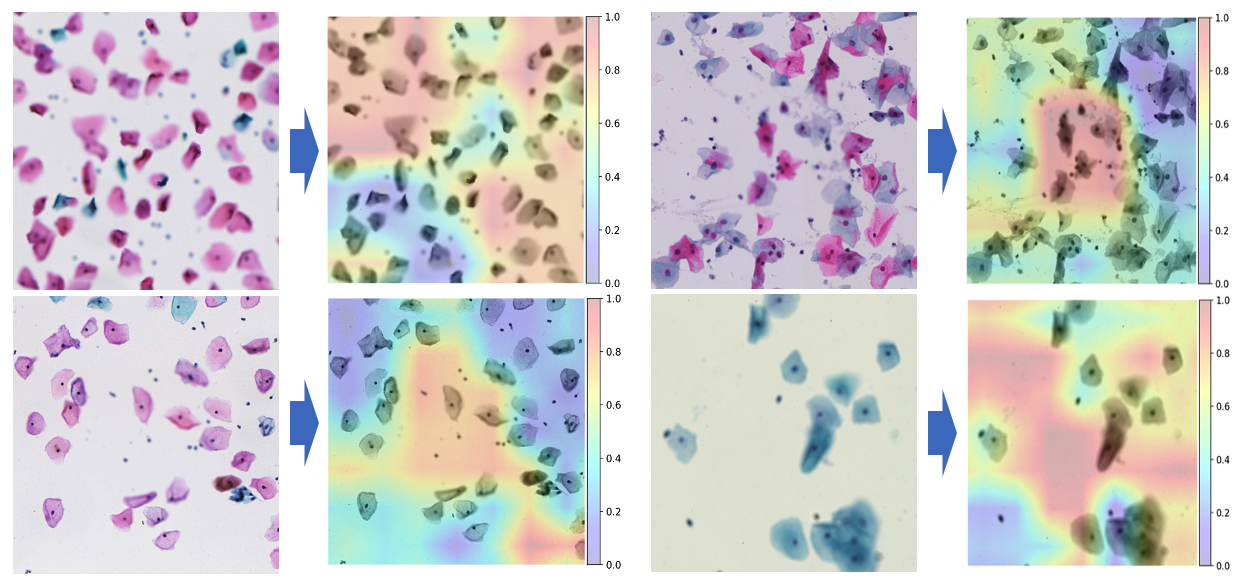}
    \caption{Visualization of FocusAttNet Model Performance in Sharpness Assessment.}
    \label{Fig.7}
    \end{figure}

 We have visualized the model outputs using heatmaps, as shown in \hyperref[Fig.7]{\figurename~\ref{Fig.7}}. The quality scores obtained from densely sampled patches are mapped to colors and overlaid on grayscale images. Scores are normalized to the $\left[ 0,1 \right]$ range before color mapping to reflect relative blurriness.
    \par
\begin{figure}[t]
    \centering
    \includegraphics[width=0.9\linewidth]{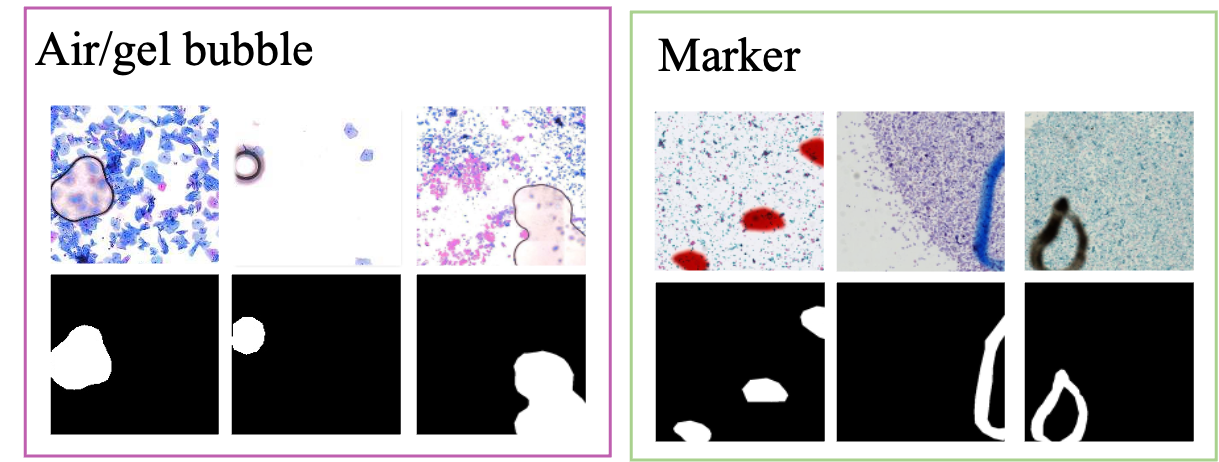}
    \caption{Segmentation Results of Marker and Air/Gel Bubble.}
    \label{Fig.8}
    \end{figure}

 In our study, we found that using lower magnification levels for quality issues such as streaks and air/gel bubbles in Whole Slide Imaging (WSI) data not only achieves good segmentation results but also significantly enhances metric calculation efficiency. For quantitative analysis, we employed WSI data at different magnifications and evaluated model performance by computing the average assessment time and average Intersection over Union (IoU). The experimental results, shown in \hyperref[table:4]{\tablename~\ref{table:4}}, indicate that at a 4x magnification, the model achieves the best balance between segmentation accuracy and efficiency, optimizing both evaluation efficiency and accuracy. We utilized the trained DoubleUNet model for segmentation, with the segmentation results illustrated in \hyperref[Fig.8]{\figurename~\ref{Fig.8}}.

    \begin{table}[h] 
    \caption{
    \label{table:4}Comparison of Different Magnification Levels.} 
    \begin{tabular*}{\tblwidth}{lll}
    \toprule \multirow{2}{*}{Magnification} & \multirow{2}{*}{\makecell{Average Evaluation \\Time/s}} & \multirow{2}{*}{\makecell{Average Test \\IOU/\%}}\\
    ~ & ~ & ~\\
    \midrule
    20x  &	29.2146	 &	98.43\\
    8x	 &	2.8706	 &	99.11\\
    4x	 &	1.2283	 &	99.08\\
    2x	 &	0.7176	 &	98.80\\
    \bottomrule 
    \end{tabular*}
    \end{table}
    
    \par To determine the optimal staining intensity range, we use cluster analysis on 200 standard samples provided by pathologist. Since the staining sample characteristics follow a normal distribution, clustering analysis identified that the standard range for the average gray value is 185 to 190 for the eosin channel and 180 to 200 for the hematoxylin channel.
    \par
\begin{figure}[h]
    \centering
    \includegraphics[width=0.9\linewidth]{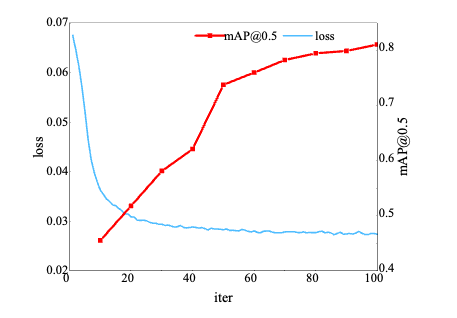}
    \caption{YOLOv5 Model Training Loss and mAP Values.}
    \label{Fig.9}
    \end{figure}

 We annotated 9,493 images using the LabelImg tool and divided the dataset into training, validation, and test sets at a ratio of 8:2:2. Through a series of experimental comparative analyses, we found that the YOLOv5 model exhibited superior performance in detecting squamous epithelial cells, cell masses, and neutrophils, achieving the highest precision, recall, and ${mAP0.5}$ metrics. Based on a comprehensive evaluation of model performance, we chose YOLOv5 as our preferred model. The trained loss and mAP values for the YOLOv5 model are shown in \hyperref[Fig.9]{\figurename~\ref{Fig.9}}. YOLOv5 demonstrated exceptional performance across multiple evaluation metrics, including detection accuracy, processing speed, and practical applicability. We present the experimental results in \hyperref[table:5]{\tablename~\ref{table:5}}.

    \begin{table*}[t] 
    \caption{
    \label{table:6}Comparison of Different Magnification Levels.} 
    \begin{tabular*}{\tblwidth}{cccccc}
    \toprule \multirow{2}{*}{WSI nubmer} & \multirow{2}{*}{\makecell{Grid-like imaging\\evaluation/s}} & \multirow{2}{*}{\makecell{Grid-like  imaging \\evaluation/s}} & \multirow{2}{*}{\makecell{Marker, air/gel bubble \\evaluation/s}} & \multirow{2}{*}{\makecell{Object  detection\\ evaluation/s}} & \multirow{2}{*}{\makecell{Staining  standard \\evaluation/s}}\\
    ~ & ~ & ~ & ~ & ~ & ~\\
    \midrule
    WSI1	 & 1.7166 	 & 119.629	 & 3.56	 & 273.9106	 & 5.8793\\
    WSI2	 & 2.2411	 & 88.604	 & 2.57	 & 244.9079	 & 5.9512\\
    WSI3	 & 2.8133	 & 112.608	 & 3.14	 & 169.5136	 & 6.4025\\
    WSI4	 & 3.3379	 & 114.257	 & 3.02	 & 331.0713	 & 7.1329\\
    WSI5	 & 1.1920	 & 57.307	 & 2.42	 & 85.7451	 & 5.5310\\
    WSI6	 & 2.3841	 & 66.640	 & 2.73	 & 151.8634	 & 5.9347\\
    Average	 & 2.1403	 & 83.274	 & 2.76	 & 209.5022	 & 5.63\\
    \bottomrule 
    \end{tabular*}
    \end{table*}

    \begin{small}
    \begin{table}[h] 
    \caption{
    \label{table:5}Comparison of Object Detection Models.} 
    \begin{tabular*}{\tblwidth}{llll}
    \toprule  \multirow{2}{*}{Model} & \multicolumn{3}{c}{dataset*}\\ 
    \cline{2-4}
    ~ & Precision & Recall & mAP@.5\\
    \midrule 
    Yolov5 \citep{redmonYouOnlyLook2016b}  & 0.7733 & 	0.7825 &  0.8107\\
    FasterRCNN \citep{renFasterRcnnRealtime2015a} & 0.7674	 & 0.7766	 & 0.7620\\
    SSD \citep{liu2016ssd} & 0.7585 & 0.7751 & 0.7322\\
    \bottomrule 
    \end{tabular*}
    \end{table}
    \end{small}

    \par To validate the effectiveness of our proposed WSI evaluation method, we conducted a comprehensive analysis on six representative WSI samples. The time expenditure for the evaluation process is summarized in \hyperref[table:6]{\tablename~\ref{table:6}}. The evaluation includes key steps: grid-like imaging quality assessment, focus evaluation, marker and air/gel bubble detection, object detection, and staining standardization. Due to the complexity of high-resolution pathological images, the overall process is time-intensive. Specifically, focus evaluation and object detection require detailed analysis, constituting most of the evaluation time. This underscores the significance of precision in pathological image analysis and its effect on the total evaluation duration.
    \par
\begin{figure}[t]
    \centering
    \includegraphics[width=0.9\linewidth]{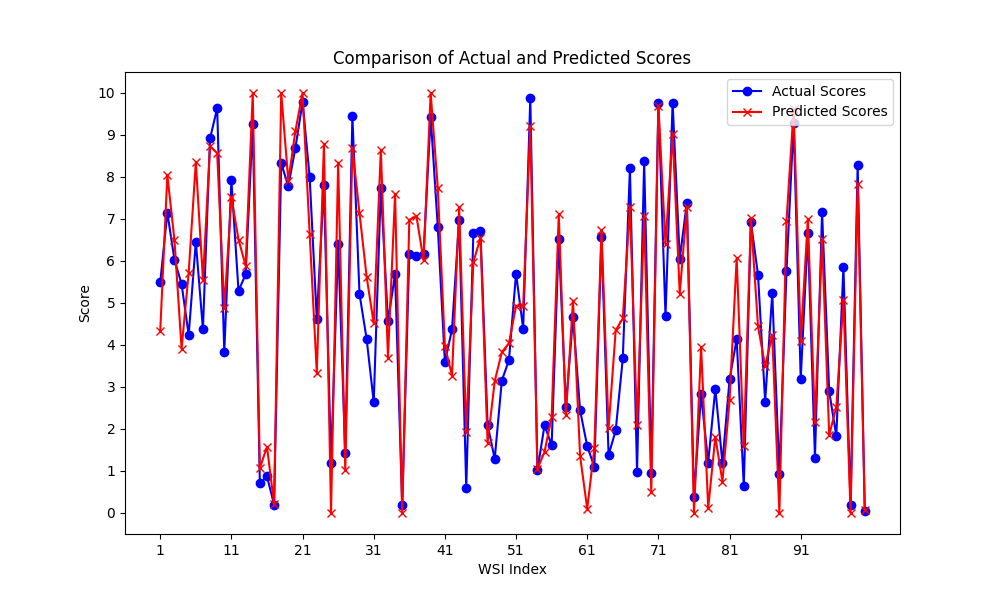}
    \caption{Comparison between True and Predicted Scores for WSI Image Quality.}
    \label{Fig.10}
    \end{figure}

 To validate the proposed evaluation method, we compared its results with those of pathologists on 100 WSI assessment tasks. The consistent analysis between the actual scores provided by pathologists and the predicted scores is shown in \hyperref[Fig.10]{\figurename~\ref{Fig.10}}. The results demonstrate that our method accurately assesses cervical cytopathology image quality, offering notable advantages in terms of speed, accuracy, and comprehensive coverage.

\section{Conclusion}
    This study addresses the limitations of current pathological image quality assessment methods, particularly in the evaluation of cervical cytopathology images. We propose a comprehensive assessment system for cervical cytopathology WSIs. Existing methods predominantly focus on singular metrics, lacking a holistic evaluation and in-depth analysis of image content, resulting in suboptimal accuracy and reliability in quality assessment.
    
    \par Our method is based on TBS diagnostic standards and pathologists' evaluation tendencies and integrates multiple deep-learning models. It utilizes multi-dimensional quality indicators such as cell count, Staining standard, sharpness, cell mass, and marker artifacts to establish a comprehensive quantitative assessment system. By learning from expert evaluation strategies, the system provides an overall WSI score and evaluates its suitability for downstream intelligent pathological diagnosis tasks.
    \par Our experiments utilized a cervical cytopathology image dataset containing 302 image samples, covering eight common quality issues. The experimental results indicate that our method outperforms traditional manual assessment methods in both speed and accuracy, delivering a more comprehensive and reliable quality evaluation. By scoring individual quality indicators and the overall quality of WSIs, we can identify and rectify quality issues originating from different sources, such as preparation and scanning processes, thereby improving the accuracy and reliability of pathological diagnoses. High-quality WSIs are archived for future research and testing while low-scoring WSIs are improved through re-preparation or re-scanning and subsequently archived.
    \par In conclusion, this study provides an efficient and precise automated solution for assessing the quality of cervical cytopathology images, significantly enhancing the reliability and practicality of pathological diagnosis. Future work will aim to optimize the model further to enhance its adaptability and performance across various scenarios.

\section*{CRediT authorship contribution statement}
Lanlan Kang: Writing - original draft, Visualization, Software, Methodology, Conceptualization, Data curation. Jian Wang: Conceptualization, Formal analysis, Writing – review and editing. Jian Qin: Writing – review \& editing, Validation, Supervision. Yiqin Liang: Formal analysis, Validation, Investigation.  Yongjun He: Conceptualization, Formal analysis, Supervision.

\section*{Declaration of competing interest}
The authors declare that they have no known competing financial interests or personal relationships that could have appeared to influence the work reported in this paper.

\section*{Data availability}
Data will be made available on request.

\section*{Acknowledgments}
Funding: This work was supported by the National Natural Science Foundation of China under Grant 62473111.


\bibliographystyle{unsrt}
\bibliography{cas-dc-template}
\end{sloppypar}
\end{document}